\documentstyle[epsf,pre,aps]{revtex}
\tighten

\def\be{\begin{equation}} 
\def\bea{\begin{eqnarray}} 
\def\ee{\end{equation}}
\def\eea{\end{eqnarray}}

\def\na{\nabla}

\def\rarrow{\rightarrow}
\def\larrow{\leftarrow}
\def\iff{\infty}

\begin{document}
\twocolumn[\hsize\textwidth\columnwidth\hsize\csname @twocolumnfalse\endcsname
\title{ Multifractal scaling of electronic transmission resonances
in perfect and imperfect Fibonacci $\delta$-function potentials.
}
\author{Prabhat K Thakur\cite{pkt} and
Parthapratim Biswas\cite{ppb}
}
\address{S.N. Bose National Centre For Basic Sciences\\
Salt Lake City, Block JD, Sector III\\
Calcutta-700 091, INDIA\\
}
\maketitle
\begin{abstract}

We present here a detailed multifractal scaling study 
for the electronic transmission resonances with the system 
size for an infinitely large one dimensional perfect 
and imperfect quasiperiodic
system represented by a sequence of $\delta$-function potentials.
The electronic transmission resonances in the energy minibands 
manifest more and more fragmented nature of the transmittance with
the change of system sizes. We claim that when a small perturbation 
is randomly present at a few number of lattice sites, the nature of 
electronic states will change and this can be understood by studying 
the electronic transmittance with the change of system size. We
report the different critical states manifested in the size 
variation of the transmittance 
corresponding to the resonant energies for both perfect and imperfect 
cases through multifractal scaling study for  few of these 
resonances.

\end{abstract}
\pacs PPACS NO. 71.25.-s, 61.44.+p, 71.90.+9
\vskip 1.5cm
]


\section{INTRODUCTION}

The discovery of quasiperiodic phase in Al-Mn alloys 
by Sehechman {\em et. al} \cite{Shech} led to a great interest 
in understanding the 
generic properties of such systems from electronic structure,
transport and  phonon dynamics point of view.  The simplest possible 
choice being to study an one dimensional quasiperiodic system.  Merlin 
and co-workers \cite{Merlin,Todd} first studied the Fibonacci 
superlattices where 
the unusual fractal like structure in the spectral properties giving 
rise to a characteristic features in experiments. Thus the 
theoretical understanding of such one dimensional quasiperiodic 
systems and the recent advances of molecular-beam-epitaxy (MBE) make these 
systems ideal candidates to study and to carry out experiments on transport
properties \cite{Todd} of these systems. Electronic 
states of one dimensional Fibonacci quasiperiodic systems have been
attempted in the recent past 
\cite{Shech,Merlin,Todd,Koho1,Koho2,Suto,Niu,Chak,Koho3,Macia2,Macia3} to 
explore some global characteristics in them.
It is well known that for such one dimensional quasiperiodic 
systems the spectrum is a 
cantor set and has self-similar structure  which results from the hierarchy
of Fibonacci lattices \cite{Koho1,Koho2,Suto,Niu,Chak,Macia1,Koho3,Macia2}. 
Most of these previous works
have mainly concentrated on studying critical nature of energy bands 
as well as the nature of wave function from localization point of view 
for isolated chains.
But less attention has been given to the 
study of the transmittance in detail to look for 
localization and delocalization aspects through some rigorous scaling 
analysis \cite{Macia2,Macia3,Thak2,Thak3}. Apart from this, most 
of the calculations have been performed starting 
from the tight-binding Anderson Hamiltonian where one models quasiperiodicity either through
the diagonal on-site energy or through the off-diagonal 
hopping term. However, the 
actual quasiperiodic potential in the Schr$\ddot{o}$dinger's equation may 
not be possible to decouple into a diagonal term which will only 
retain quasiperiodicity  
and a constant hopping term and vice versa. 
So one should expect that some of the physical features 
one finds in the numerical investigation through the continous 
models may explain more closely the experimental situation.
It is possible to fabricate semiconductor
heterostructure with a variety of potential 
profiles along the growth directions, such as 
rectangular and parabolic ones in resonant 
tunneling devices and saw-tooth potential in delta 
doped layers \cite{Jacos}. Apart from that, it has been realized in the 
recent past that in some of the thin
quasi one dimensional wires, the variation of 
the potentials at the impurities is much larger 
than that of the energies involved by the weak electric field; 
the Kronig-Penny model is then the 
natural choice for a relevant model to start with.

 In recent years, many authors have 
attempted the problem of studying 
the scattering of free electrons 
by an array of $\delta$-function potentials
to investigate the electronic states and 
transport properties of one dimensional 
disordered systems, quasiperiodic sequences, incommensurate 
systems \cite{Azbel1,Azbel2,Sokol,Avis,Far,Thak1}
and in particular the transport properties 
of quasiperiodic superlattices\cite{Jacos}.
Here we have attempted the study of size dependence of 
transmittance through the multifractal scaling analysis. 
We have focussed our 
attention in characterising the 
transmittance resonances in the 
different energy minibands that one observes
in an array of one dimensional Fibonacci $\delta$-function potentials. 
This has been done in the same spirit where in a disordered chain 
one studies the transmittance vs energy plot having signature of 
exponentially localized states along with a few stochastic resonances 
also known as Azbel resonances\cite{Azbel1,Azbel2}.

The strong variation of transmittance 
with system size gives an indication of 
the non-trivial nature of transmittance 
for the various resonant energies.  
Based on a detailed numerical investigation of 
the fluctuating  pattern of the transmittance 
within the framework of multifractal scaling of 
transmittance with system size, it is possible to 
infer on the delocalization and localization aspects 
for both perfect and imperfect one 
dimensional quasiperiodic systems respectively.
In this work we will see that both the perfect and imperfect 
realisations resemble some novel size
variation which is intermediate between the so called exponential decay 
and the homogeneous Bl\"och like contributions. 
We claim that many of these are resonant states are of {\em nearly critical}
nature instead of showing either exponential decay or 
homogeneous spread along the chain. 
We have  made a rigorous  study for the
electronic transmittance in imperfect Fibonacci 
system to verify whether a minor 
change in the potential in a perfect chain can destroy the critical nature 
of the states altogether leading to a complete localization
of electronic states. Recently, some works in the area of
imperfect quasiperiodic chains have been reported \cite{Nau} claiming 
the instability of the spectrum of a Fibonacci superlattice due to 
introduction of a single defect. 
We have seen that  for many of the localized states the strong spatial 
fluctuations in the transmittance will exhibit nearly self-similar
features like critical states upto a large length scale  
followed by a long-tailed oscillatory pattern. However, 
many of the resonant critical states due to the 
presence of imperfectness also
change but show some fluctuations reflecting some 
kind of  instability 
in the $\alpha$-$f(\alpha)$ spectrum subject to large size variation.

\section {Brief outline of the model}

 Let us consider the dynamics of electrons in a
Fibonnaci quasiperiodic array 
of $\delta$-function potentials.
We know that a Fibonacci sequence $S_j$ of 
order $j$ is obtained by $j$ successive 
applications of the transformations : 

\bea
S_{j+1} &\larrow & S_j \oplus S_{j-1} \\
A       &\rarrow & AB \nonumber \\
B       &\rarrow & A \nonumber
\eea
Where $S_{0} = A $ and $S_{1} = AB$ and the symbol 
$\oplus $ means the operation of composition. 
For a very large but finite $j$ one  can generate 
an infinitely large array of $\delta$-function 
potentials in the  Fibonnaci sequence where 
the following relations hold: 
\bea
  F_{j+1} = F_j+F_{j-1}\quad \quad \mbox{and} \\
  \lim_{j \to \iff} \left(\frac{F_{j-1}}{F_j}\right) = \left(\frac{{\sqrt 5}-1}{2}\right)
\eea
$F_{j} $ being the number of atoms in the sequence $S_j$.
Recently S$\acute{a}$nchez {\em et. al.}\cite{Macia1}  have made 
an explicit derivation 
of the generalized Poincare map and calculated the electronic transmittance 
due to scattering from a lattice with Fibonacci \cite{Macia1} and 
correlated  disorder 
potentials within the Kronig-Penny model\cite{Macia1}. 
In this paper we have based our numerical computations 
of electronic transmittance using the same 
Poincare map analysis having a very simple 
and straight forward mathematical structure. 
This algorithm has been numerically 
verified to be very stable and suitable 
for doing calculation for very large 
system size in one dimensional systems. 
We illustrate here only a brief outline 
of the model. 
A detailed derivation of the generalized Poincare map 
{\em etc.} will be found in the ref.\cite{Macia1}.
We concentrate on the Fibonacci model 
for an electron interacting with lattice potentials of the following form.
\bea
V(x) = \sum_{n=1}^{N} \Gamma_{n}\,\delta(x-x_{n}) 
\eea
 We have chosen the position of the $\delta$-function potentials to be 
regularly placed ({\it i.e} $x_n=na, a=1$) with the strength $\Gamma_A$ and $\Gamma_B$ for $A$ and $B$ 
type of atom respectively.
The Schr$\ddot{o}$dinger's equation  
is then given by ( using $ \hbar^2 = 2m = 1 $), 
\bea
\left[ -\na^2 + \sum_{n=1}^{N}\Gamma_n \,\delta(x-x_n) \right]\Psi(x) = E\Psi(x)
\label{kn:eq1}
\eea
One can now introduce reflection and transmission amplitude $R_N, T_N$  
respectively through the following relationship:
\bea
     \Psi(x) &  = & e^{ikx} + R_N e^{-ikx}  \quad \quad \mbox{ $x < 1$ } \\
     \Psi(x) &  = & T_N e^{ikx}  \qquad \qquad \quad \mbox{\,\,\, $x \ge N$ } 
\eea
The tranmission amplitude 
$T_N$ after $N$ sites is given by the recurrence relation,
\bea
A_N = (\alpha_N +\frac{\alpha_{N-1}^*\beta_N}{\beta_{N-1}})A_{N-1}
-(\frac{\beta_N}{\beta_{N-1}})A_{N-2} 
\label{kn:eq2}
\eea
where  
\bea
          A_N & = & \frac{1}{T_N^*} \\
     \alpha_N & = & \left( 1- i\frac{\Gamma_N}{2k}\right)e^{ik} \\
     \beta_N  & = & -i\left(\frac{1}{2k}\right)\Gamma_N e^{-ik}
\eea

The initial conditions $A_0=1$ and $A_1= \alpha_1$ are sufficient to determine 
the transmission amplitude $T_N $ 
completely for an arbitrary size {\it i.e} for a 
of large number of $\delta$-function potentials. 
The transmittance $t_N$  is given by 
\bea
   t_N = T_N^{*}T_N 
       = \left(\frac{1}{A_N}\right)\left(\frac{1}{A_N}\right)^*
\label{kn:eq6}
\eea
\section{Multifractal scaling analysis for the 
variation of transmittance with system size}
 In the recent past, Thakur {\em et. al.} made 
some electron localization
studies in an aperiodic potential where the variation of transmittance 
with the system size has been investigated\cite{Thak2}. 
It has been established that this variation 
of the tranamittance with the increase of system size in an  
infinitely large system exhibits some intrinsic 
global features which can be fully 
captured by the multifractal scaling analysis\cite{Halsey}.
The same scaling aspects have also been used 
for the Azbel resonance problem \cite{Azbel2}
in one dimensional random chain as well as
in the study of extended and the critical 
states in the vicinity of dimer resonance 
energy in the random dimer problem\cite{Thak3}. 
Following this line of approach, 
our motivation in the present work is to examine the
electronic transmission of different resonant characters seen in the 
transmittance versus energy diagram 
of a quasiperiodic Fibonacci superlattices. 
We observe that it is possible to explain the dependence of 
the transmittance with the system size through the multifractal scaling 
aspects of the transmittance.  
Here we outline the mathematical content of the multifractal scaling for 
electronic transmittance 
only because of its direct connection with our computation 
and its relation to some of the typical transport features manifested 
through some global properties in the different energy minibands
in our calculations. 
 We will discuss the same in the context of the multifractal 
scaling algorithm 
developed by Chabbra and Jensen 
in brief\cite{Chabbra}. This algorithm has been successfully 
used in analyzing the crossover states in generalized 
Aubry model, random dimer model and also in many other 
contexts in the literature\cite{Thak2,Thak3,Sch}. 
The multifractal formalism  has been developed essentially
to describe the statistical properties of some measure in 
terms of its distribution 
of the singularity spectrum $f(\alpha)$ corresponding to its 
singularity strength $\alpha$. In our calculation we take  
the normalized transmittance $P_i$ as the required measure which is given by 
$$
  P_{i} = \frac{T_{i}}{\sum_{i=1}^{N} T_{i}} \nonumber
$$
where $T_{i} $ is the transmittance from one 
end of the system upto the $i$-th segment 
when the length of the chain is divided 
into $N$ equal small segments such that 
the transmittance for a given size  is 
obtained by always increasing the previous size 
by adding the same number of atoms. According 
to Chabbra and Jensen if we define 
the $q$-th moment of the probability measure $P_{i}$ by
the following expression:
$$
  \mu_{i}(q,N) = \frac{P_i^{q}}{\sum_{i=1}^{N} P_i^{q}}
$$
then a complete characterization of the fractal singularities 
can be made in terms of the function $\mu_i(q, N)$. 
Hence one can derive the spectrum of 
fractal singularities defined by, $(\alpha,f(\alpha))$. 
The expression for  $f(\alpha)$ 
is given by,  
\be
f(\alpha) = \lim_{N \to \iff} -\frac{1}{\log N}\sum_{i=1}^{N} \mu_{i}(q,N)\,\log \,\mu_i(q,N) 
\ee 
and the corresponding singularity strength $\alpha$  is obtained as 
\be
\alpha = \lim_{N \to \iff}  -\frac{1}{\log N}\sum_{i=1}^{N} \mu_{i}\,\log\, P_i 
\ee

In this regard, the specific nature of electronic transmittance, 
reflected through 
$\alpha_{max}$ {\em i.e} large negative $q$, $\alpha_{min}$ {\em i.e} 
for large positive $q$, and also through $f(\alpha_{max})$ and 
$f(\alpha_{min})$ values, 
can be inferred by the following characteristics for large $N( N\to \iff)$ :
\begin{enumerate}

\item 

Extended nature : $\alpha_{min} \to 1 $, $f(\alpha_{min}) \to 1 $, 
                    $\alpha_{max} \to 1 $, $f(\alpha_{max}) \to 1 $.

\item 
Localized nature : $\alpha_{min} \to 0 $, $f(\alpha_{min}) \to 0 $, 
                    $\alpha_{max} \to \iff $, $f(\alpha_{max}) \to 1$.

\item 
Critical nature : $\alpha $ versus $f(\alpha)$ curve will show a tendency 
                  to converge onto a single curve as $N \to \iff$. 

Ideally for extended nature the $\alpha$ versus $f(\alpha)$ 
will contract eventually to $(1,1)$ whereas for critical nature 
the $\alpha$ versus $f(\alpha)$ curve, with the increase of system size, 
may sometimes show minor fluctuations as well.

\end{enumerate}
\section {Results and discussion}
In this section, we discuss the results for the electronic transmittance 
of both the perfect and the imperfect 
Fibonacci superlattices through solving the Eq.(\ref{kn:eq1}) making use of the 
mathematical relations Eqs.(\ref{kn:eq2}-\ref{kn:eq6}).
Here we mainly focus our attention on the electronic transmission 
resonances in different energy minibands and have investigated  
whether the resonances survive when 
we increase the system size drastically by plotting the transmittance 
versus energy plots for various system sizes. 
As stated earlier, our aim is to examine 
whether the delocalized characteristics of the resonances show some 
interesting universal properties for such quasiperiodic potentials even 
when subject to a large size variation.  We finally claim that our numerical 
studies of multifractal scaling for the variation of transmittance 
with system size capture most of the generic features of electronic 
states and transport in the Fibonacci quasiperiodic superlattice 
within the Kronig-Penny model. 

\subsection{Results for a perfect Fibonacci sequence }

 In this subsection we discuss the results for a 
perfect Fibonacci quasiperiodic potential. In Fig.1 we have 
shown the transmittance $T$ versus $E$ plots for $\Gamma_A= 2$ units and
$\Gamma_B=1 $ units for three different system sizes, $N= 75025 $
 (lower curve), $N=196418$ (middle curve), and $N=317811$(topmost curve).
One can clearly see that as the system size 
increases transmission envelope fragments more and more. It's also 
true that at any stage, in general, the $T$ versus $E$ plot has a 
fragmented shape. These characteristic features are due to the 
critical nature of the energy spectrum of the Fibonacci 
quasiperiodic system. We now choose arbitrarily a transmission resonance 
corresponding to a small region of energy minibands. This resonance 
can be identified more clearly if the $T$ versus $E$ plot within this 
regime can be checked for a very fine energy mesh $\sim 10^{-7}$ or 
even less such that a resonanance within this fine energy resolution 
will exhibit it's signature in the $T$ versus $N$ plot.
In Fig.2(a) we have chosen $E= 20.0918599999603$ units as one of such resonant 
energies and have shown the variation of the transmittance $T$ with system 
size $N$. The $T$ versus $N$ plot clearly shows highly oscillatory 
pattern having a very complicated structure and even for variation of size 
from $10^5$ to $7 \times 10^5$  no. of atoms or layers it retains 
the same global shape in 
it's pattern. We have carried out the multifractal scaling of the 
transmittance with it's system size to analyze this spatial pattern 
in Fig.2(c). Also in Fig.2(b) we have shown another plot of $T$ versus $N$ 
for $2 \times 10^5$ to $3 \times 10^5$ no. of atoms or layers for a resonant 
energy $E=3.204406000$ units. We see that the oscillatory pattern in Fig.2(b)
seems to be different from that of Fig.2(a) altogether. In Fig.2(c) 
the $\alpha-f(\alpha)$ plot shows a small inward contraction with the 
increase of system size from $N=196418$ to $N=317811$ no. of atoms or layers 
and then from $N=317811$
to $N=514229$ no. of atoms or layers. The deviation of $\alpha_{min}$ and 
$\alpha_{max}$ values with the size variation is much less and 
is $ \sim 0.01$. We identify them by saying {\em nearly critical 
resonant states}.
In Fig.2(d) the $\alpha$ versus $f(\alpha) $ plot corresponding to 
the variation of the transmittance$T$ with system size $N$ in Fig.2(b),
 have been shown. Here we see that as the system size has been increased 
from $N=75025$ no. of atoms or layers to $N=196418$ no. of atoms or layers the curve 
deviates from the initial position as a whole but when the system size 
has been increased to $N=317811$ no. of atoms or layers, the curve comes in between the two 
previous positions. However, here the amount of deviation of $\alpha_{max}$ 
values are $\sim 0.025 $ to $ 0.040$ and $\alpha_{min} \sim 0.01$  to 
$0.02$. One knows that for a localized state with the increase of systen size 
$\alpha_{max} $ values systematically goes out but here for the size 
$N=317811$ it comes inside. This is the signature of critical state  
showing that it's neither localized nor extended but have self-similar
pattern for the size variation upto an infinitely large system size. 
We identify this resonant state as {\em critical resonant state} . 
We now consider 
set of parameters for the potentials, namely, $\Gamma_A=3.0$ units and
$\Gamma_B=1.0$ units and check whether here also one finds some typical resonant
states. In Fig.3(a), the $T$ versus $N$ plot has been shown for the 
resonant energy $E=6.07038530468009$ units. The variation of the transmittance ($T$)
with system size ($N$) manifests a highly fragmented and oscillatory 
behavior from $5\times 10^5$ to $ 832040$ no. of atoms or layers. 
We also choose another resonant energy $ E = 15.0532885305118$ units for the same potential 
parameters $\Gamma_A $ and $\Gamma_B$ where the $T$ versus $N$ plot shows 
clearly some oscillatory as well as fragmented 
pattern. However, here, the overall shape 
approxomately gives a periodic oscillation.
The $f(\alpha)$ versus $\alpha$ plot corresponding to the variation of 
transmittance($T$) with system size($N$) in Fig.3(a) has been shown 
in Fig.3(c). Note that as we increase the system size from $N=75025$ 
to $N=317811$ both the $\alpha_{max} $ and $\alpha_{min}$ values change
by very small amount $\sim 0.01$ or less. As we increase the size to $N=514229$
, the curve begins to deviate outwards but when we increase the size from 
$N=514229$ to $N=832040$ no. of atoms or layers the curve then again shows 
an inward contraction.
So we see here that the qualitative behavior is much like a 
resonant state in Fig.2(c) and 
this can be identified as a critical resonant state as well.
Next we analyze the $\alpha$ versus $f(\alpha)$ plot
in Fig.3(d) for the resonant state corresponding to the  
energy $E= 15.0532885305118$ units  for which the variation of 
transmittance with system size has 
been shown in Fig.3(b). Here we notice that, as we increase the system size 
from $N= 196418$ to $N= 317811$ no. of atoms or layers the curve undergoes an 
inward contraction. Next with the subsequent increase of system sizes
upto $N=1346269$ no. of atoms or layers, the curves systematically contracts 
as we increase the system size more and more, the curves show a relative 
tendency  to contract by very lesser amount. We identify this to be 
qualitatively similar to the resonant states as depicted in Fig.2(c). 
So we again identify it to be a {\em nearly critical resonant state} and
it is in some sense different from the critical resonant shown in Fig.2(b)
and Fig.3(c).

 We think these two typical resonant states, namely, critical and 
 nearly critical can be identified for different 
choice of potentials also. It's also true 
that they are of two distinct nature which one can differentiate through 
the $\alpha$ versus $f(\alpha)$ plots. 
These resonant states exist for different energy 
regimes and also for any choice of  
$\Gamma_A$ and $\Gamma_B$. This particular aspect is true since 
we have chosen the resonant energies in Fig.2(a) and Fig.2(b) or in Fig.3(a) 
and Fig.3(b) in completely different energy regimes  
arbitrarily.

\subsection{Results for an imperfect Fibonacci sequence }

 Now we go over to analyze the effects of disorder present 
at random in any layer. The amount of perturbation in 
the potential parameters $\Gamma_A $ and $\Gamma_B$ 
have been considered $5\%$ at random in any layers through 
perturbing 2$\%$ of the atoms or layers in the system as a whole. 

 In Fig.4(a) we have shown the transmittance($T$) 
versus system size($N$) 
plot for the resonant energy $E = 6.07062311828030$ units
corresponding to starting Fibonacci potentials 
$\Gamma_A=3.0$ units and $\Gamma_B=1.0 $ 
unit and then a fractional change in the potntials 
$f$= 0.05 have been considered 
randomly at any layer where  2$\%$ of the layers 
have been perturbed as a whole. One can see
that the transmittance amplitudes look like different 
patches of spikes of many different heights. 
The variation of transmittance with system size($N$) 
has been shown from $10^6$ to $1346269$ no. of
layers. Note that it looks neither like
 extended nature or exponentially localized 
but appranetly the profile of the spikes seem to have some power law decay. 
We have checked through the multifractal scaling analysis whether the resonant states 
would show some systematic decay or not. 

In Fig.4(b), the plot of $f(\alpha)$ versus $\alpha$ 
corresponding to the variation of 
transmittance with system size in Fig.4(a), have been shown. 
We have shown $\alpha$ versus $f(\alpha)$ 
curves corresponding to the size variation starting with the initial size $832040$ no. of atoms or layers
and then succesively increasing no. of atoms or layers by $10^5$ no. of atoms or layers. 
One can clearly see in Fig.4(b) that 
$\alpha$ versus $f(\alpha)$ curves show some instability even 
for the variation of transmittance beyond
$10^6$ no. of atoms or layers. This instability in the $\alpha$ versus $f(\alpha)$ plot is due to the effect 
of disorder in the Fibonacci quasiperiodic layers. However, if 
one looks at the overall variation 
of the transmittance with system size, from $10^6$ to $1346269$ no. 
atoms or layers then it shows the decay 
but of vary complicated form and this is also 
captured in the outermost $\alpha$ versus $f(\alpha)$ 
plot corresponding to the variation of transmittance $T$ 
for the system of size from $832040$ to $1346269$ 
no. of atoms or layers eventually. One can see that this is the typical 
long-tailed fluctuations which we have observed 
due to the interplay of the effect of disorder 
and the resonant nature of critical state in the ideal system. 
Since  this is a state 
which shows a size variation in between critical and exponentially  
localized state, one can identify this resonant state, due to 
the presence of imperfectness in the system, 
to be more closely like {\em critical resonant state}. 
Next we have shown, in Fig.5(a), the variation of transmittance 
with system size for a resonant state corresponding to 
the energy $E= 4.3987600006124$ units
and for the values $\Gamma_A=2.0$ units,$\Gamma_B=1.0$ units. 
It's a typical localized state with fluctuations 
throughout but having less fluctuations in the tail part 
as compared to that in the beginning,
$i.e$, around $50,000$ to $2 \times 10^5$ no. of atoms or layers.
In Fig.5(b), the $\alpha$ versus $f(\alpha)$ plots corresponding to
the variation of transmittance with system size as shown in Fig.5(a), for 
 $N = 317811, 514229$ and, $ N = 832040 $ 
no. of atoms or layers, have been shown. It's quite clear that with the increase of 
system size the curves begin to move outwards reflecting 
the very nature of a localized state having an overall systematic decay.
In Fig.5(c) we have shown 
the plot of transmittance with the system size $N$ for an 
imperfect superlattice with $E=20.88552401746 
$ units. This is also a resonant state with strong fluctuations 
throughout which arises because of the presence of disorder in the 
system. 
In Fig.5(d) we have shown the $\alpha$ versus $f(\alpha)$
plot corresponding to the variation of the transmittance($T$) 
with system size ($N$)in Fig.5(c). Note that, as the system size has been 
increased from $ N = 196418$ to $317811$ no. of atoms or layers, 
the curves undergo an outward shift but as we increase the system size 
from $317811$ to $514229$ no. of atoms or layers it again goes in between 
the two previous positions. This sort of instability reflected in the 
$\alpha$ versus $f(\alpha)$ is due to the presence of imperfectness 
coupled with the long-ranged quasiperiodic order in the system.
Since these states do not show any systematic decay either in the 
$\alpha - f(\alpha)$ plot, and show some kind of critical nature,
we identify them more closely like {\em critical resonant states} also.
Therefore, to paraphrase, the manifestation 
of the effect of disorder on 
the resonant critical states, apart from the existence of having 
systematic decay, has been 
found to be twofold: firstly, it has long-tailed  
fluctuations and secondly having sustained strong fluctuations 
even when subject to 
large size variations as mentioned above.

\section{conclusion}

In this work we explore the nature of electronic states 
and transport in one dimensional $\delta$-functional Fibonacci potentials
from a rigorous study of electronic transmission resonances corresponding
to the different energy minibands. In this regard, although some 
studies of transmittance have been reported in the past, our numerical 
study of the size dependence of the same is directly associated with the 
characterization of the nature of electronic states and transmittance using the multifractal
scaling aspects which to our knowledge has not yet
been investigated adequately till date.
We should point out the fact that many of the resonances show strong 
oscillations with the variation of the system size and the oscillating pattern 
continues to survive till the width of the minibands, in which it belongs, 
goes to zero and an energy gap opens up there. In a perfect Fibonacci case, 
critical behavior of the transmittance with system size has its origin from 
the typical nature of the Fibonacci potential and at least a few of them can be 
distinguished from the others. More specifically, here, the clear distinction 
between the different critical resonant states have been possible with the help of 
multifractal scaling analysis.
On the other hand, we have considered the imperfect 
situation to be a case where the imperfectness
appears randomly as a homogeneous  substitutional disorder in 
the different lattice points causing only minor changes 
in the Fibonacci potentials which may be present $e.g$ in an experimental
realisation of Fibonacci superlattice.

In the imperfect case we have examined that this 
may induce localization effects which may be manifested 
through some non-exponential spatial decay of transmittance and 
the resonant states in the perfect situation will appear
again more closely like critical resonant states but with some 
modified nature.

 The long-tailed oscillatory pattern of the transmittance shows a very slow 
spatial decay and it retains its nearly self-similar pattern 
even if one drastically changes the system sizes. Although the 
overall pattern of the oscillations eventually 
show decaying fluctuations, the decay is not uniform$-$ 
indicating an interplay between the long-ranged quasiperiodic order  
and the disorder an electron experiences with the increase of system size. 
The other manifestation of the interplay between 
long-ranged quasiperiodic order and substitutional disorder is the instability 
of the resonant state in the variation of transmittance 
with the increase of system size. This has been also reflected in 
the $\alpha$-$f(\alpha)$ spectrum in our numerical 
studies.  We, therefore, hope that it 
will be interesting to search for both non-exponential 
localization and critical properties of the
electronic transmittance through designing some controlled experiments in 
this direction. 
We believe that our work will shed some more light 
in understanding the theoretical notion of the novel scaling 
properties for the electron localization and 
delocalization aspects directly in the size dependence 
of the conductance in Fibonacci quasiperiodic systems. 

\section*{ACKNOWLEDGEMENTS}
 Parthapratim Biswas would like to thank the Council of Scientific 
 and Industrial Research (CSIR) for financial assistance in the 
 form of a senior research fellowship.



\begin{figure}
\caption{ 
Tranamittance $T$ versus energy $E$ 
plots for $\Gamma_A = 2$  units and $\Gamma_B = 1$ units
. $N$ = 75025 no. of atoms or layers (lower curve , $N$ = 196418 
no. of atoms or layers (middle curve), $N$= 317811 no. of atoms or layers (topmost curve).
}
\end{figure}


\begin{figure}
\caption{ 
(a) Transmittance $T$ versus system size $N$ for a pure Fibonacci lattice with 
$\Gamma_A$ = 2 units, $\Gamma_B= 1$ unit  and $E = 20.0918599999603$  units. \\ \\
(b) Transmittance $T$ versus system size $N$ for a pure Fibonacci lattice with 
$\Gamma_A = 2$ units, $\Gamma_B= 1$ unit  and $E = 3.20440600$ units. \\ \\ 
(c) $f(\alpha)$ versus $\alpha$ plot for the same set of parameters 
as in Fig. 2(a) but for different system sizes having maximum no. of atoms or layers
$N= 196418$, (line), $N= 317811$
(dashed line) and for $N=514229$ (smaller dashed line), 
where in each case  starting size is $5$ no. of atoms or layers and the rest is 
obtained increasing $5$ no. of atoms or layers each time.
\\ \\
(d)
$f(\alpha)$ versus $\alpha$ plot for the same set of parameters 
as in Fig.2(b) but for different system sizes having maximum no. of atoms or layers
$N= 75025$, (line), $N= 196418$
(dashed line) and for $N=317811$ (smaller dashed line),
where in each case  starting size is $5$ no. of atoms or layers and the rest is 
obtained increasing $5$ no. of atoms or layers each time.
}
\end{figure}


\begin{figure}
\caption{ 
(a) Transmittance $T$ versus system size $N$ for a pure Fibonacci lattice with 
$\Gamma_A = 3$ units , $\Gamma_B= 1$ unit  and $E = 6.07038530468009$ units. \\ \\
(b) Transmittance $T$ versus system size $N$ for a pure Fibonacci lattice for the
same set of potential parameters as in Fig.3(a) but for energy
value $E = 15.0532885305118$ units. \\ \\
(c) $f(\alpha)$ versus $\alpha$ plot for the same set of parameters 
as in Fig.3(a) for different system sizes having maximum no. of atoms or layers
$N= 75025$ (line), $N= 317811$ (dashed line)
, $N=514229$ (smaller dashed line) and  $N= 832040$(dotted line), 
where in each case  starting size is $5$ no. of atoms or layers and the rest is 
obtained increasing $5$ no. of atoms or layers each time. \\ \\
(d)
$f(\alpha)$ versus $\alpha$ plot for the same set of parameters 
as in Fig.3(b) for different system sizes having maximum no. of atoms or layers
$N= 196418$ (line), $N= 317811$(dashed line)
, $N=514229$ (smaller dashed line), $N= 832040$(dotted line) and 
for $N= 1346269$(dash-dotted line),
where in each case  starting size is $5$ no. of atoms or layers and the rest is 
obtained increasing $5$ no. of atoms or layers each time.
}
\end{figure}


\begin{figure}
\caption{ 
(a) Transmittance $T$ versus system size $N$ for an imperfect  
Fibonacci lattice for $\Gamma_A = 3.0$ units ,$\Gamma_B= 1$ unit
for $E = 6.07062311828030$ units. The fractional change $f
= 0.05$ has been considered randomly at any layer or atom (say $i$)
so that the new $ \Gamma_i=\Gamma_i+ f*\Gamma_i $. The concentration
of such defects present in the lattice is $c = 0.02$
\\ \\
(b) $f(\alpha)$ versus $\alpha$ plot for the same set of parameters 
as in Fig.4(a) for different system sizes. The data set 
have been considered starting with size having no. of atoms or layers
$N=832040$ and adding $10^5$ no. of atoms or layers (line), adding $2 \times 10^5$
no. of atoms or layers(dashed line), $3 \times 10^5$ no. of atoms or layers (smaller 
dashed line), $4 \times 10^5$ no. of atoms or layers (dotted line) and  $5 
\times 10^5$ no. of 
layers (dash-dotted line)  with steps of $5$ layers in each case.
}
\end{figure}

\begin{figure}
\caption{ 
(a)Transmittance $T$ versus system size $N$ for an imperfect  
Fibonacci lattice for the set of parameters 
$\Gamma_A = 2.0$  units  and $\Gamma_B = 1.0 $ unit  
for $E = 4.39876000006124$ units . The fractional change $f
= 0.05$ has been considered randomly at any layer or atom 
so that the new $ \Gamma_i=\Gamma_i+ f*\Gamma_i$. The concentration
of such defects present in the lattice is $c = 0.02$. \\ \\
(b)
$f(\alpha)$ versus $\alpha$ plot for the same set of parameters 
as in Fig.5(a) for different system sizes. The data set 
have been considered starting with size having maximum no. of atoms or layers
$N= 317811$ (line), $N= 514229$ (dashed line.) and 
$N = 832040$(smaller dashed line) no. of atoms or layers,
where in each case  starting size is $5$ no. of atoms or layers and the rest is 
obtained increasing $5$ no. of atoms or layers each time.
\\ \\
(c) Transmittance $T$ versus system size $N$ for an imperfect  
Fibonacci lattice for the same set of potential parameters as in Fig.5(a)
but for $E = 20.88552401746$ units.  \\ \\
(d) $f(\alpha)$ versus $\alpha$ plot for the same set of parameters 
as in Fig.5(c) for different system sizes. The data set 
have been considered starting with size having maximum no. of atoms or layers
$N=196418$ (line), $N= 317811$ (dashed line) and $N= 514229$ (smaller 
dashed line.),
where in each case  starting size is $5$ no. of atoms or layers and the rest is 
obtained increasing $5$ no. of atoms or layers each time.
}
\end{figure}
\end{document}